\documentclass[pre,aps,12pt]{revtex4-1}
\usepackage{graphicx,color}

\begin{document}

\title{Thermodynamics of delta-chain with ferro- and antiferromagnetic
interactions}
\author{D.~V.~Dmitriev}
\author{V.~Ya.~Krivnov}
\email{krivnov@deom.chph.ras.ru} \affiliation{Institute of
Biochemical Physics of RAS, Kosygin str. 4, 119334, Moscow,
Russia.}
\author{J.~Richter}
\affiliation{Institut f\"{u}r Physik,
          Otto-von-Guericke-Universit\"{a}t Magdeburg,
          P.O. Box 4120, 39016 Magdeburg, Germany}
\affiliation{Max-Planck-Institut f\"{u}r Physik komplexer Systeme,
          N\"{o}thnitzer Stra\ss e 38, 01187 Dresden, Germany}
\author{J.~Schnack}
\affiliation{Fakult\"at f\"ur Physik, Universit\"at Bielefeld, Postfach
100131, D-33501 Bielefeld, Germany}
\date{}

\begin{abstract}
Motivated by a novel cyclic compound $Fe_{10}Gd_{10}$ with record
ground state spin in which the arrangement of magnetic ions with
$s=\frac{5}{2}$ and $s=\frac{7}{2}$ corresponds to a saw-tooth
chain we investigate the thermodynamics of the delta-chain with
competing ferro- and antiferromagnetic interactions. We study both
classical and quantum versions of the model. The classical model
is exactly solved and quantum effects are studied using full
diagonalization and a finite temperature Lanczos technique for
finite delta-chains as well as modified spin wave theory. It is
shown that the main features of the magnetic susceptibility of the
quantum spin delta chain are correctly described by the classical
spin model, while quantum effects significantly change the
low-temperature behavior of the specific heat. The relation of the
obtained results to the $Fe_{10}Gd_{10}$ system is discussed.
\end{abstract}

\maketitle

\section{Introduction}

\begin{figure}[tbp]
\includegraphics[width=5in,angle=0]{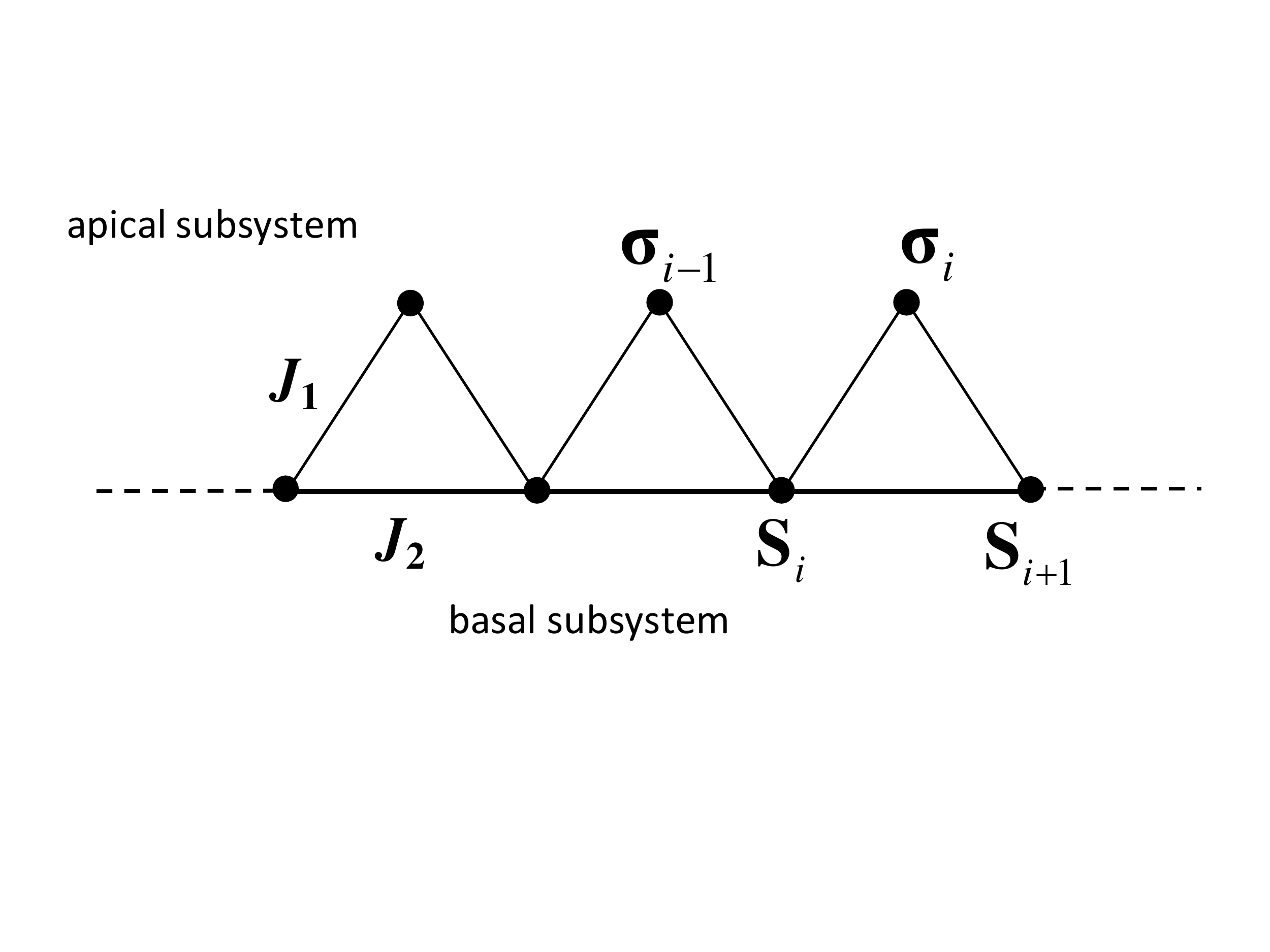}
\caption{The delta-chain model.}
\label{Fig_saw}
\end{figure}

Low-dimensional quantum magnets on geometrically frustrated
lattices have attracted much interest in last years \cite{diep,mila,qm}. An
important class of such systems includes lattices consisting of
triangles. An interesting and typical example of these objects
is the delta or the sawtooth Heisenberg model consisting of a
linear chain of triangles as shown in Fig.~\ref{Fig_saw}. The
Hamiltonian of this model has the form:
\begin{equation}
H=J_{1}\sum_{i=1}^{N}\mathbf{\sigma }_{i}\cdot (\mathbf{S}_{i}+\mathbf{S}%
_{i+1})+J_{2}\sum_{i=1}^{N}\mathbf{S}_{i}\cdot \mathbf{S}_{i+1}
\label{Hq}
\end{equation}

The interaction $J_{1}$ acts between the apical ($\mathbf{\sigma
}_{i}$) and the basal ($\mathbf{S}_{i}$) spins, while $J_{2}$ is
the interaction between neighboring basal spins. A direct
interaction between the apical spins is absent. The quantum
$s=\frac{1}{2}$ delta-chain with antiferromagnetic (AF) exchange
interactions $J_{1}$ and $J_{2}$ ($J_{1},J_{2}>0$) has been
studied extensively and it exhibits a variety of peculiar
properties \cite{flat,Zhit,Schmidt,Honecker,Derzhko2004}. At the
same time the $s=\frac{1}{2}$ delta-chain with ferromagnetic
$J_{2}$ and antiferromagnetic $J_{1}$ interaction (F-AF
delta-chain) is very interesting as well and has unusual
properties. In particular, the ground state of this model is
ferromagnetic for $a =\frac{J_{2}}{\left\vert J_{1}\right\vert
}<\frac{1}{2}$ and as it is believed \cite{Tonegawa} that it is
ferrimagnetic for $a >\frac{1}{2}$. The critical point $a
=\frac{1}{2}$ is the transition point between these two ground
state phases. The $s=\frac{1}{2}$ F-AF delta-chain at the critical
point $a =\frac{1}{2}$ has been studied in Ref.~\cite{KDNDR}. It
is an example of a quantum system with a flat excitation band
\cite{flat,Flach}, that provides the possibility to find several
rigorous results for the quantum many-body system at hand. Thus,
it was shown \cite{KDNDR} that the ground state at the critical
point consists of localized multi-magnon complexes, and it is
macroscopically degenerate. An additional motivation for the study
of this model is the existence of real compounds, malonate-bridged
copper complexes \cite{Inagaki,Tonegawa,ruiz,Kaburagi}, which are
described by this model.

The $s=\frac{1}{2}$ F-AF model can be extended to the delta-chain
composed of two types of spins ($\mathbf{\sigma
}_{i},\mathbf{S}_{i}$) characterized by the spin quantum numbers
$S_{a}$ and $S_{b}$ of the apical and basal spins, respectively.
The ground state of this model is ferromagnetic for $a
=\frac{J_{2}}{\left\vert J_{1}\right\vert }<\frac{S_{a}}{2S_{b}}$
and non-collinear ferrimagnetic for $a
>\frac{S_{a}}{2S_{b}}$. The critical point between these phases is
$a _{c}=\frac{S_{a}}{2S_{b}}$. The ground state at the critical
point consists of exact multi-magnon states as well as for the
$s=\frac{1}{2}$ model and has similar macroscopic degeneracy
\cite{KDNDR}.

Recently a mixed $3d/4f$ cyclic coordination cluster \\
$[Fe_{10}Gd_{10}(Me-tea)_{10}(Me-teaH)_{10}(NO_{3})_{10}]20MeCN$
(i.e. $Fe_{10}Gd_{10}$) has been synthesized and studied
\cite{S60}. This cluster consists of $10+10$ alternating
gadolinium and iron ions and its spin arrangement corresponds to
the delta chain with $Gd$ and $Fe$ ions as the apical and basal
spins correspondingly. As it was established in Ref.~\cite{S60}
the exchange interaction between $Fe$ ions is antiferromagnetic
($J_{2}=0.65K$) and the interaction between $Fe$ and $Gd$ is
ferromagnetic ($J_{1}=-1.0K$). The spin values of $Fe$ and $Gd$
ions are $S=\frac{5}{2}$ for $Fe^{III}$ and $S=\frac{7}{2}$ for
$Gd^{II}$. The ground state spin of this cluster is $S=60$ which
is one of the largest spin of a single molecule. This molecule is
a finite-size realization of the F-AF delta-chain with
$S_{a}=\frac{7}{2}$ and $S_{b}=\frac{5}{2}$. Remarkably, according
to the estimate of the values of $J_1$ and $J_2$ in
Ref.~\cite{S60} the frustration parameter is $a =0.65$, i.e. it is
very close to the critical value of $a _{c}=0.7$. Therefore, this
molecule, although it is not directly at the critical point and
located in the F phase, has properties which are strongly
influenced by the nearby quantum critical point. Because the spin
quantum numbers for $Fe$ and $Gd$ ions are rather large it seems
that the classical approximation for ($S_{a},S_{b}$) F-AF
delta-chain is justified, except at very low temperatures when
quantum fluctuations can substantially change the properties of
the system.

To obtain the classical version of Hamiltonian (\ref{Hq}) we set
$\sigma _{i}=S_{a}\vec{n}_{i}$ and $S_{i}=S_{b}\vec{n}_{i}$, where
$\vec{n}_{i}$ is the unit vector at the i-th site. Taking the
limit of infinite $S_{a}$ and $S_{b}$ with a finite ratio
$\frac{S_{a}}{S_{b}}$ we arrive at the Hamiltonian of the
classical delta chain
\begin{equation}
H=-\sum_{i=1}^{2N}\vec{n}_{i}\cdot \vec{n}_{i+1}+\alpha %
\sum_{i=1}^{N}\vec{n}_{2i-1}\cdot \vec{n}_{2i+1},  \label{H}
\end{equation}%
where $N$ is the number of triangles. In (\ref{H}) we take the
apical-basal interaction as $-1$ and the basal-basal interaction
$\alpha$ as
\begin{equation}
\alpha =\frac{J_{2}S_{b}}{\left\vert J_{1}\right\vert S_{a}}=a\frac{S_{b}}{%
S_{a}}  \label{a}
\end{equation}
which is the frustration parameter of the model.

The ground state phase diagram of the classical model consists of
a ferromagnetic phase at $a<\frac{S_{a}}{2S_{b}}$ ($\alpha
<\frac{1}{2}$) and a ferrimagnetic one at $a>\frac{S_{a}}{2S_{b}}$
($\alpha >\frac{1}{2}$). Remarkably, the transition between
these phases occurs at the same frustration parameter
($a_{c}=\frac{S_{a}}{2S_{b}}$) as in the quantum model. In terms
of $\alpha $ the critical point between the ferromagnetic and
ferrimagnetic phases is at $\alpha =\frac{1}{2}$.

One of the goals of this paper is the study of the thermodynamics
of the classical version of F-AF Heisenberg model (\ref{Hq}),
where we put special attention on the parameter region
corresponding to $Fe_{10}Gd_{10}$. In what follows we use the
normalized temperature
\begin{equation}
t=\frac{T}{\left\vert J_{1}\right\vert S_{a}S_{b}}  \label{beta}
\end{equation}%
and the corresponding inverse temperature $\beta =1/t$
to present  the thermodynamic properties of model (\ref{H}).
Since the classical ground state exhibits a non-trivial macroscopic
degeneracy for $\alpha
> \frac{1}{2}$, see Sec.~II, we may expect unconventional low-temperature
physics especially in the ferromagnetic regime close to the critical point  $\alpha_c=\frac{1}{2}$.

The effect of quantum fluctuations at low temperatures will be
studied by a combination of full exact diagonalization (ED) using
J. Schulenburg's {\it spinpack} code
 \cite{spinpack}
 and
the finite temperature Lanczos (FTL) technique \cite{FTL1,FTL2} as
well as by the modified spin-wave theory (MSWT) \cite{MSWT}.

The paper is organized as follows. In Sec.~II we describe the
ground state of model (\ref{H}) in different regions of
frustration parameter $\alpha $. The partition function, the
correlation functions, the specific heat and magnetic
susceptibility are calculated in Sec.~III. In Sec.~IV explicit
analytical results for the partition function, the spin
correlation functions and the magnetic susceptibility in the
low-temperature limit are presented for different regions of the
parameter $\alpha $. In Sec.~V the scaling law near the critical
point $\alpha =\frac{1}{2}$ is established and finite-size effects
are estimated. In Sec.~VI the quantum effects in the ferromagnetic
phase are studied with a particular focus on that value of the
frustration parameter $\alpha $ which is relevant for
$Fe_{10}Gd_{10}$.

\section{Ground state}

We start our study of model (\ref{H}) from the determination of
the ground state. For this aim it is useful to represent
Hamiltonian (\ref{H}) as a sum over triangle Hamiltonians
\begin{equation}
H=\sum_{i=1}^{N}H_{\Delta }(i) , \label{Hsum}
\end{equation}%
where the Hamiltonian of a triangle has the form
\begin{equation}
H_{\Delta }(i)=-\vec{n}_{2i-1}\cdot \vec{n}_{2i}-\vec{n}_{2i}\cdot \vec{n}%
_{2i+1}+\alpha \vec{n}_{2i-1}\cdot \vec{n}_{2i+1}  \label{Htri} .
\end{equation}

To determine the ground state of model (\ref{Hsum}) we need to
find the spin configuration on each triangle which minimizes the
classical energy. It turns out that the lowest spin configuration
on a triangle is different in the regions $\alpha \leq \frac{1}{2}$
and $\alpha >\frac{1}{2}$. For $\alpha \leq \frac{1}{2}$ the
ground state is the trivial ferromagnetic one with all spins on each
triangle pointing in the same direction. The global spin
configuration of the whole system in this case is obviously
ferromagnetic as well.

\begin{figure}[tbp]
\includegraphics[width=4in,angle=0]{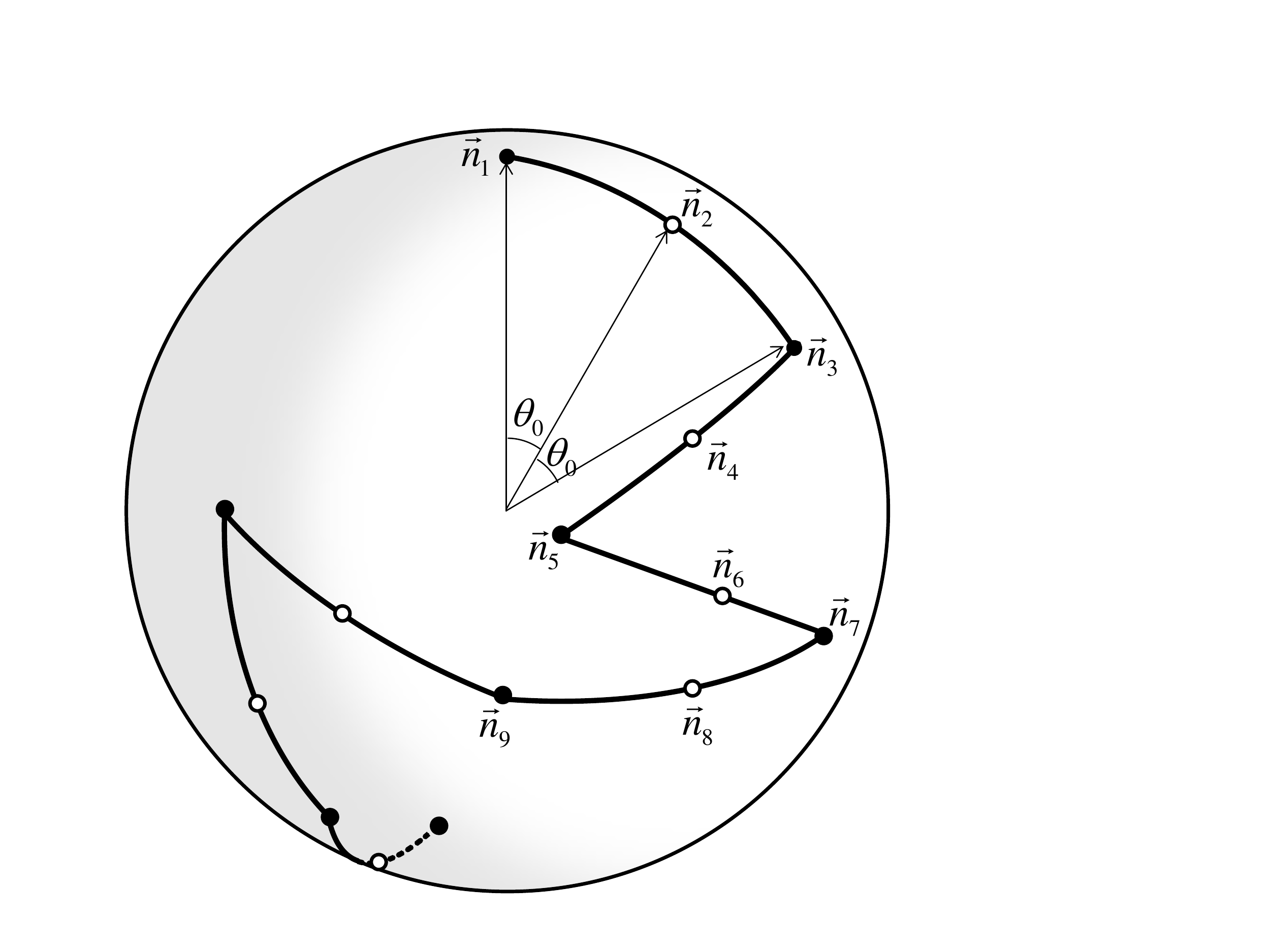}
\caption{Ground state spin configurations of model (\ref{H}) for
$\alpha >\frac{1}{2}$ as random walk on unit sphere.}
\label{Fig_sphere}
\end{figure}

For $\alpha >\frac{1}{2}$ the lowest classical energy on each
triangle is given by the ferrimagnetic configuration, where all
spins of triangle $\vec{n}_{1},\vec{n}_{2},\vec{n}_{3}$ lie in the
same plane and the spin $\vec{n}_{2}$ forms an equal angle $\theta _{0}$
with spins $\vec{n}_{1}$ and $\vec{n}_{3}$:
\begin{eqnarray}
\vec{n}_{1}\cdot \vec{n}_{2} &=&\vec{n}_{2}\cdot \vec{n}_{3}=\cos \theta _{0}
\nonumber \\
\vec{n}_{1}\cdot \vec{n}_{3} &=&\cos \left( 2\theta _{0}\right)  \nonumber \\
\cos \theta _{0} &=&\frac{1}{2\alpha }  \label{theta0}.
\end{eqnarray}
So, each triangle has the non-zero magnetization $m_{\Delta
}=1/\alpha +1$, the direction of which coincides with the apical
spin $\vec{n}_{2}$. The spins of the next triangle
($\vec{n}_{3},\vec{n}_{4},\vec{n}_{5}$) also form the
ferrimagnetic configuration in the ground state. But in the
general case the spins ($\vec{n}_{3},\vec{n}_{4},\vec{n}_{5}$) can
lie in any plane, formed by the rotation of the first triangle
plane around the spin $\vec{n}_{3}$ by an arbitrary angle
\cite{Chandra}. So, the ground state of the second triangle is
degenerate over the angle between planes
($\vec{n}_{1},\vec{n}_{2},\vec{n}_{3}$) and
($\vec{n}_{3},\vec{n}_{4},\vec{n}_{5}$). Then the plane of the
third triangle ($\vec{n}_{5},\vec{n}_{6},\vec{n}_{7}$) is rotated
by an arbitrary angle around spin $\vec{n}_{5}$, and so on. Hence,
the global ground state of the whole system for $\alpha
>\frac{1}{2}$ is macroscopically degenerate.
Each ground state
spin configuration for $\alpha >\frac{1}{2}$ can be represented as
a sequence of the points lying on the unit sphere with an equal
distance between the neighboring points as shown in
Fig.~\ref{Fig_sphere}.

\section{Partition function}

The partition function $Z$ of model (\ref{H}) is%
\begin{equation}
Z=(\prod_{i=1}^{2N}\int d\Omega _{i})\exp \left( -\beta H\right) ,
\label{Z}
\end{equation}%
where $d\Omega _{i}$ is the differential of the solid angle for
the $i$-th spin, $d\Omega _{i}=\sin \theta _{i}d\theta
_{i}d\varphi _{i}/4\pi $.
Using the dual transformation employed in Ref.~\cite{Harada} we
choose a local coordinate system connected with the $i$-th spin.
The $z_{i}$ axis is parallel to $\vec{n}_{i\text{ }}$and the
$y_{i}$ axis is in the plane spanned by $\vec{n}_{i}$ and
$\vec{n}_{i+1}$. A new set of the angles $(\theta _{i},\varphi
_{i})$ is introduced, where $\theta _{i}$ is the angle between
$\vec{n}_{i+1}$ and $\vec{n}_{i}$ and $\varphi _{i}$ is the angle
between the components of $\vec{n}_{i+1}$ and $\vec{n}_{i-1}$
projected onto the ($x_{i}$,$y_{i}$) plane.
In terms of these variables the Hamiltonian on the triangle
(\ref{Htri}) becomes
\begin{equation}
H_{\Delta }(i)=-\cos \theta _{2i-1}-\cos \theta _{2i}+\alpha \cos
\theta _{2i-1}\cos \theta _{2i}+\alpha \sin \theta _{2i-1}\sin
\theta _{2i}\cos \varphi _{2i}  \label{Htri1} .
\end{equation}
As follows from the latter equation, the total Hamiltonian
(\ref{Hsum}) does not contain angles $\varphi _{1},\varphi
_{3},\varphi _{5}\ldots $, on which the partition function can be
integrated. As a result, the partition function reduces to the
product of independent multipliers and takes the form
\begin{equation}
Z=\prod_{i=1}^{N}Z_{i}=Z_{\Delta }^{N} , \label{Z2}
\end{equation}%
where $Z_{\Delta }$ is `the partition function' of an isolated triangle%
\begin{equation}
Z_{\Delta }=\frac{1}{8\pi }\int_{0}^{\pi }\sin \theta _{1}d\theta
_{1}\int_{0}^{\pi }\sin \theta _{2}d\theta _{2}\int_{0}^{2\pi
}d\varphi _{2}e^{-\beta H_{\Delta}(1)} . \label{Z3}
\end{equation}
The integral over the angle $\varphi _{2}$ can be carried out analytically:%
\begin{equation}
Z_{\Delta }=\frac{1}{4}\int_{0}^{\pi }\sin \theta _{1}d\theta
_{1}\int_{0}^{\pi }\sin \theta _{2}d\theta _{2}e^{-\beta H_{1}}I_{0}\left(
\beta H_{2}\right) , \label{ZA}
\end{equation}%
where%
\begin{eqnarray}
H_{1} &=&-\cos \theta _{1}-\cos \theta _{2}+\alpha \cos \theta
_{1}\cos \theta _{2}  \nonumber \\
H_{2} &=&\alpha \sin \theta _{1}\sin \theta _{2}  \label{H1H2}
\end{eqnarray}%
and
\begin{equation}
I_{0}\left( x\right) =\frac{1}{2\pi }\int_{0}^{2\pi }e^{x\cos \varphi
}d\varphi  \label{Bessel}
\end{equation}%
is the Bessel function of imaginary argument.

Thus, the problem of the calculation of the partition function of model (\ref%
{H}) is reduced to the double integral (\ref{ZA}). All thermodynamic
quantities can be expressed through the corresponding derivatives of the
partition function.

\subsection{Specific heat}

The specific heat can be expressed through the second derivative
of the partition function. However, in order to avoid loss of
accuracy caused by the numerical derivatives it is convenient to
use the following equation for the calculation of the specific
heat:
\begin{equation}
C=\beta ^{2}\left\langle H_{\Delta }^{2}\right\rangle -\beta
^{2}\left\langle H_{\Delta }\right\rangle ^{2} . \label{c}
\end{equation}
Here $\left\langle H_{\Delta }\right\rangle $ is the energy of each triangle
(\ref{Htri}), which is
\begin{equation}
\left\langle H_{\Delta }\right\rangle =-2\left\langle \vec{n}_{1}\cdot
\vec{n}_{2}\right\rangle +\alpha \left\langle \vec{n}_{1}\cdot \vec{n}_{3}
\right\rangle,
\end{equation}
where local correlators $\left\langle \vec{n}_{1}\cdot
\vec{n}_{2}\right\rangle $ and $\left\langle \vec{n}_{1}\cdot
\vec{n}_{3}\right\rangle $ on one triangle are given by
\begin{eqnarray}
\left\langle \vec{n}_{1}\cdot \vec{n}_{2}\right\rangle &=&\frac{1}{8\pi
Z_{\Delta }}\int_{0}^{\pi }\sin \theta _{1}d\theta _{1}\int_{0}^{\pi }\sin
\theta _{2}d\theta _{2}\int_{0}^{2\pi }d\varphi _{2}e^{-\beta H_{\Delta
}}\cos \theta _{1}  \label{n1n2} \\
\left\langle \vec{n}_{1}\cdot \vec{n}_{3}\right\rangle &=&\frac{1}{8\pi
Z_{\Delta }}\int_{0}^{\pi }\sin \theta _{1}d\theta _{1}\int_{0}^{\pi }\sin
\theta _{2}d\theta _{2}\int_{0}^{2\pi }d\varphi _{2}e^{-\beta H_{\Delta
}}\left( \cos \theta _{1}\cos \theta _{2}+\sin \theta _{1}\sin \theta
_{2}\cos \varphi _{2}\right) . \nonumber 
\end{eqnarray}
The expectation value $\left\langle H_{\Delta }^{2}\right\rangle $
is
\begin{equation}
\left\langle H_{\Delta }^{2}\right\rangle =\frac{1}{4Z_{\Delta }}%
\int_{0}^{\pi }\sin \theta _{1}d\theta _{1}\int_{0}^{\pi }\sin \theta
_{2}d\theta _{2}F\left( \theta _{1},\theta _{2}\right) e^{-\beta H_{\Delta }},
\label{H^2}
\end{equation}%
where%
\begin{equation}
F\left( \theta _{1},\theta _{2}\right) =\left( H_{1}^{2}+H_{2}^{2}\right)
I_{0}\left( \beta H_{2}\right) +H_{2}\left( 2H_{1}+\beta ^{-1}\right)
I_{1}\left( \beta H_{2}\right)  \label{F}
\end{equation}%
and $H_{1}$, $H_{2}$ are given by Eqs.~(\ref{H1H2}).

\begin{figure}[tbp]
\includegraphics[width=5in,angle=0]{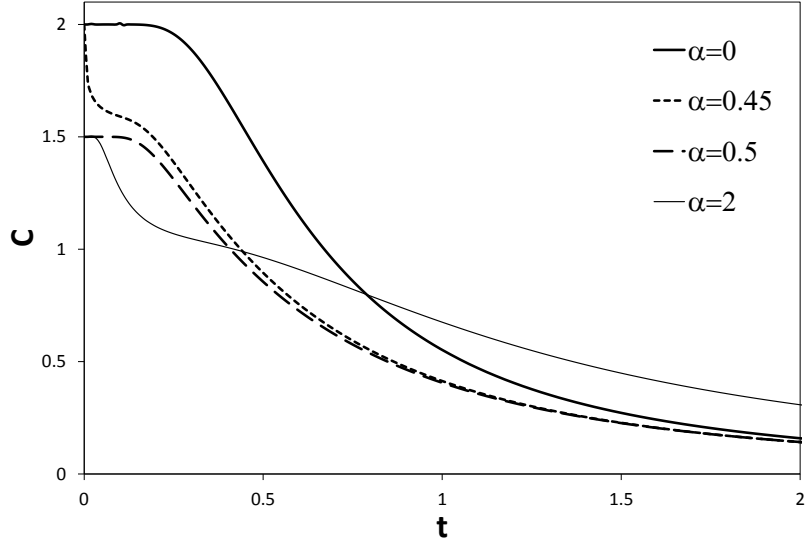}
\caption{Specific heat as a function of the normalized temperature
for the model (\ref{H}) for several values of $\alpha $.}
\label{Fig_CT}
\end{figure}

In Fig.~\ref{Fig_CT} we present the results for the specific heat
per triangle as a function of the normalized temperature $t$ for
different values of $\alpha $. Note that $\alpha=0.45 $
corresponds to the situation in $Fe_{10}Gd_{10}$.  As it is seen
in Fig.~\ref{Fig_CT} $C(t)$ (in $k_B$ units) has different low
temperature limits for $\alpha <\frac{1}{2}$ and $\alpha \geq
\frac{1}{2}$:
\begin{eqnarray}
C(0) &=&2,\qquad \alpha <\frac{1}{2}  \nonumber \\
C(0) &=&\frac{3}{2},\qquad \alpha \geq \frac{1}{2}.
\end{eqnarray}
$C(t)$ approaches $2$ in the ferromagnetic region $\alpha
<\frac{1}{2} $ because the model is the classical one with two
degrees of freedom per spin (four degrees per triangle). But for
$\alpha \geq \frac{1}{2}$ the specific heat tends to
$\frac{3}{2}$. It means that only three degrees of freedom per
triangle contribute to the specific heat and there is one local
rotational degree ($\varphi_3$) of freedom which costs no energy
and gives no contribution to the thermodynamics.

The low temperature behavior of $C(t)$ for $\alpha $ slightly
lower the point $\frac{1}{2}$ has the following characteristic
feature. $C(t)$ sharply increases from $C\simeq \frac{3}{2}$ to
$C=2$ when $t$ tends to zero. As will be shown in Sec.~V this
feature is consistent with the scaling dependence of physical
quantities in the vicinity of the critical point $\alpha
=\frac{1}{2}$. According to Eq.~(\ref{scale_x}), see below, the scaling
variable is $t/(\alpha -\frac{1}{2})^{2}$. Therefore, we can say
that the system behaves as in the critical point $\alpha
=\frac{1}{2}$ for temperatures $t>(\alpha -\frac{1}{2})^{2}$, and
starts to feel the small deviation from the critical point at very
low temperatures $t<(\alpha -\frac{1}{2})^{2}$.

\begin{figure}[tbp]
\includegraphics[width=5in,angle=0]{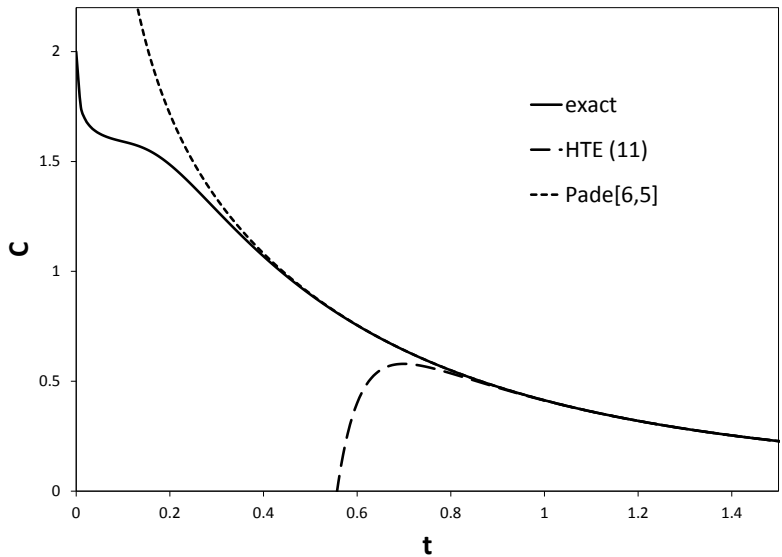}
\caption{Comparison of the specific heat calculated by 11th-order HTE
and the corresponding Pade approximant [6,5] with
the exact classical result for $\alpha=0.45$.} \label{Fig_c_HTE}
\end{figure}

For high temperatures we obtained an analytical expansion up to
11-th order in $t^{-1}$ \cite{11order}. The leading terms of this high
temperature expansion (HTE) are
\begin{equation}
C=\frac{2+\alpha ^{2}}{3t^{2}}-\frac{2\alpha}{3t^{3}}+O(t^{-4}).
\label{Ct_HTE}
\end{equation}
The results of HTE and the corresponding Pade approximant [6,5]
are compared with the exact classical $C(t)$ for $\alpha=0.45$ in
Fig.~\ref{Fig_c_HTE}. As it is seen in Fig.~\ref{Fig_c_HTE} the raw HTE
series
coincides with exact result for $t\geq 0.9$ and  the Pade
approximant [6,5] extends the confidence
region up to $t\geq 0.4$.

\subsection{Correlation functions}

In this subsection we derive the expressions for the spin
correlation function $\vec{n}_{i}\cdot \vec{n}_{j}$ and the zero
field susceptibility.
Since we work with the local coordinate system directed so that
the spin vector on each site is directed along the $z$ axis, to
find the scalar product of spin vectors $\vec{n}_{l}\cdot
\vec{n}_{l+r}$ we need to express the vector $\vec{n}_{l+r}$ in
the local coordinate system located on the site $l$. This can be
represented as a chain of successive rotations \cite{Harada}:
\begin{equation}
\vec{n}_{l}\cdot \vec{n}_{l+r}=\left( 0,0,1\right) \hat{T}_{2}\hat{T}%
_{3}\ldots \hat{T}_{r}\left(
\begin{array}{c}
0 \\
0 \\
1
\end{array}
\right) ,  \label{corrT}
\end{equation}
where
\begin{equation}
\hat{T}_{i}=\hat{R}^{x}(\theta _{i-1})\hat{R}^{z}(\varphi _{i})  \label{T}
\end{equation}%
and the rotation operators over the axes $x$ and $z$ are
\begin{eqnarray}
\hat{R}^{x}(\theta ) &=&\left(
\begin{array}{ccc}
1 & 0 & 0 \\
0 & \cos \theta & \sin \theta \\
0 & -\sin \theta & \cos \theta%
\end{array}%
\right)  \label{Rx} \\
\hat{R}^{z}(\varphi ) &=&\left(
\begin{array}{ccc}
-\cos \varphi & -\sin \varphi & 0 \\
\sin \varphi & -\cos \varphi & 0 \\
0 & 0 & 1%
\end{array}%
\right) .  \label{Rz}
\end{eqnarray}
Then, for the averages one has%
\begin{equation}
\left\langle \vec{n}_{l}\cdot \vec{n}_{l+r}\right\rangle =\frac{1}{Z_{\Delta
}^{r}}\prod_{i=1}^{r}\frac{1}{4\pi }\int_{0}^{\pi }\sin \theta _{i}d\theta
_{i}\int_{0}^{2\pi }d\varphi _{i+1}e^{-\beta H}\cdot \left( 0,0,1\right)
T_{2}T_{3}\ldots T_{r}\left(
\begin{array}{c}
0 \\
0 \\
1%
\end{array}%
\right) \, .
\end{equation}
Since total Hamiltonian (\ref{Hsum}) does not contain the angles
$\varphi _{1},\varphi _{3}\ldots $, integration over these angles
can be carried out explicitly, which results in
\begin{equation}
\int_{0}^{2\pi }\hat{R}^{z}(\varphi )\frac{d\varphi }{2\pi }=\left(
\begin{array}{ccc}
0 & 0 & 0 \\
0 & 0 & 0 \\
0 & 0 & 1%
\end{array}%
\right) =\left(
\begin{array}{c}
0 \\
0 \\
1%
\end{array}%
\right) \cdot \left( 0,0,1\right) .
\end{equation}
This implies that the long-distance correlator splits into the
product of correlators on all intermediate triangles
\begin{equation}
\left\langle \vec{n}_{1(2)}\cdot \vec{n}_{2r+1(0)}\right\rangle
=\left\langle \vec{n}_{1(2)}\cdot \vec{n}_{3}\right\rangle \left\langle \vec{%
n}_{3}\cdot \vec{n}_{5}\right\rangle \left\langle \vec{n}_{5}\cdot \vec{n}%
_{7}\right\rangle \ldots \left\langle \vec{n}_{2r-1}\cdot \vec{n}%
_{2r+1(0)}\right\rangle .
\end{equation}
Using the fact that all local correlators of type $\left\langle
\vec{n}_{2i-1}\cdot \vec{n}_{2i+1}\right\rangle $ are equal to
$\left\langle \vec{n}_{1}\cdot \vec{n}_{3}\right\rangle $ and
$\left\langle \vec{n}_{2}\cdot \vec{n}_{3}\right\rangle
=\left\langle \vec{n}_{1}\cdot \vec{n}_{2}\right\rangle $ we
obtain the expressions for the spin correlation functions:
\begin{eqnarray}
\left\langle \vec{n}_{2n+1}\cdot \vec{n}_{2m+1}\right\rangle &=&\left\langle
\vec{n}_{1}\cdot \vec{n}_{3}\right\rangle ^{\left\vert m-n\right\vert }
\nonumber \\
\left\langle \vec{n}_{2n}\cdot \vec{n}_{2m+1}\right\rangle &=&\left\langle
\vec{n}_{1}\cdot \vec{n}_{2}\right\rangle \left\langle \vec{n}_{1}\cdot \vec{%
n}_{3}\right\rangle ^{\left\vert m-n\right\vert }  \nonumber \\
\left\langle \vec{n}_{2n}\cdot \vec{n}_{2m}\right\rangle &=&\left\langle
\vec{n}_{1}\cdot \vec{n}_{2}\right\rangle ^{2}\left\langle \vec{n}_{1}\cdot
\vec{n}_{3}\right\rangle ^{\left\vert m-n\right\vert -1} . \label{corr}
\end{eqnarray}
The local correlators $\left\langle \vec{n}_{1}\cdot \vec{n}%
_{2}\right\rangle $ and $\left\langle \vec{n}_{1}\cdot \vec{n}%
_{3}\right\rangle $ are given by Eqs.~(\ref{n1n2}).

As follows from Eq.~(\ref{corr}) the spin correlation functions
decay exponentially  with the correlation length
\begin{equation}
\xi =-\frac{1}{\ln \left\vert \left\langle \vec{n}_{1}\cdot \vec{n}%
_{3}\right\rangle \right\vert } . \label{ksi}
\end{equation}
The analysis of the behavior of correlation functions (\ref{corr})
will be given in Sec.~V.

Now we can calculate the zero-field magnetic susceptibility per
spin
\begin{equation}
\chi =\frac{1}{6TN}\sum_{i,j}\left\langle \vec{S}_{i}\cdot \vec{S}%
_{j}\right\rangle . \label{chi_0}
\end{equation}
Using the obtained correlation functions (\ref{corr}) we arrive at
the following expression for the magnetic susceptibility:
\begin{equation}
\chi t=\frac{\left( 1+x\left\langle \vec{n}_{1}\cdot \vec{n}%
_{2}\right\rangle \right) ^{2}}{3x(1-\left\langle \vec{n}_{1}\cdot \vec{n}%
_{3}\right\rangle )}+\frac{x^{2}-1}{6x} , \label{chi}
\end{equation}%
where $x=S_{a}/S_{b}$.

The high temperature behavior of the susceptibility is obtained in
the analytical form up to 11-th order in $t^{-1}$ \cite{11order}. The leading
terms of HTE at $x=1$ are
\begin{equation}
\chi (t)=\frac{1}{3t}+\frac{2-\alpha }{9t^{2}}+\frac{2-4\alpha
+\alpha^2}{27t^{3}}+O(t^{-4})
\end{equation}

\begin{figure}[tbp]
\includegraphics[width=5in,angle=0]{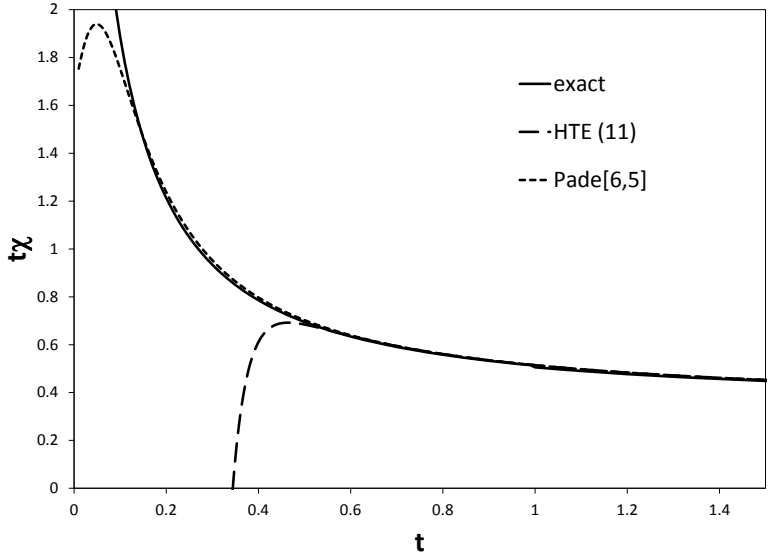}
\caption{Comparison of the susceptibility
 calculated by 11th-order HTE
and the corresponding Pade approximant [6,5]
with
the exact classical result for $\alpha=0.45$.}
\label{Fig_chiT_HTE}
\end{figure}

The dependencies $\chi(t)$ obtained by HTE up to 11-th order and
the corresponding Pade approximant [6,5] for $\alpha=0.45$ are
compared with the exact result in Fig.~\ref{Fig_chiT_HTE}. As one
can see in Fig.~\ref{Fig_chiT_HTE} the raw HTE series separates from the
exact result at $t\sim 0.5$, whereas the Pade approximant [6,5]
coincides with the exact result up to $t\sim 0.2$.

In the case of the ferromagnetic short-range order, when spins on
one triangle $\vec{n}_{1},\vec{n}_{2},\vec{n}_{3}$ are almost
parallel, the correlation length (\ref{ksi}) is large and the
susceptibility relates to the correlation length as
\begin{equation}
\chi t=\frac{\left( 1+x\right) ^{2}}{3x}\xi . \label{chi-xi}
\end{equation}

\section{Low temperature limit}

In general, Eq.~(\ref{chi}) completely describes the behavior of
the magnetic susceptibility as a function of the temperature and
the frustration parameter $\alpha $. However, in this Section we
pay special attention to the low temperature limit, where the
explicit analytical results will help us to establish the scaling
law near the critical point.

At $T\to  0$ the integration in Eq.~(\ref{Z3}) can be carried out
using the saddle point method. For this aim we need to expand the
exponent in Eq.~(\ref{Z3}) near the ground state of the triangle
Hamiltonian (\ref{Htri1}). Since the ground state of $H_{\Delta }$
is different in the regions $\alpha <\frac{1}{2}$ and $\alpha
>\frac{1}{2}$ and at the transition point $\alpha =\frac{1}{2}$,
it is necessary to study these three cases separately.

\subsection{The case $\alpha <\frac{1}{2}$}

In the case $\alpha <\frac{1}{2}$ all spins pointing in the same direction ($\theta _{i}=0$)
in the ground state of $H_{\Delta}$
and we expand the exponent in Eq.~(\ref{Z3}) near the energy
minimum, so that the partition function (\ref{Z3}) takes the form
\begin{equation}
Z_{\Delta }=\frac{1}{8\pi }e^{(2-\alpha )\beta }\int_{0}^{\infty
}\theta _{1}d\theta _{1}\int_{0}^{\infty }\theta _{2}d\theta
_{2}\int_{0}^{2\pi }d\varphi _{2}\exp \left[ -\frac{1}{2}\beta
(1-\alpha)(\theta _{1}^{2}+\theta _{2}^{2})-\beta \alpha \theta
_{1}\theta _{2}\cos \varphi _{2}\right] .
\end{equation}
Performing the integration we obtain
\begin{equation}
Z_{\Delta }=\frac{\exp \left( 2\beta -\alpha \beta \right)
}{4\beta ^{2}\left( 1-2\alpha \right) } . \label{Zz1}
\end{equation}

The correlator $\left\langle \vec{n}_{1}\cdot
\vec{n}_{2}\right\rangle =\left\langle \cos \theta
_{1}\right\rangle $ in the low-temperature limit is
\begin{equation}
\left\langle \vec{n}_{1}\cdot \vec{n}_{2}\right\rangle =1-\frac{1}{2}%
\left\langle \theta _{1}^{2}\right\rangle .
\end{equation}
We omit technical details and give the expression for the
expectation value $\left\langle \theta _{1}^{2}\right\rangle $:
\begin{equation}
\left\langle \theta _{1}^{2}\right\rangle =2t\frac{1-\alpha }{1-2%
\alpha } .  \label{theta1}
\end{equation}
The correlator $\left\langle \vec{n}_{1}\cdot
\vec{n}_{3}\right\rangle $ given by Eq.~(\ref{n1n2}) is calculated
in a similar way, which yields:
\begin{eqnarray}
\left\langle \vec{n}_{1}\cdot \vec{n}_{2}\right\rangle &=&1-t\frac{1-\alpha}{1-2\alpha }  \nonumber \\
\left\langle \vec{n}_{1}\cdot \vec{n}_{3}\right\rangle &=&1-t\frac{2}{1-2%
\alpha } . \label{corr1}
\end{eqnarray}
Now, substituting Eqs.~(\ref{corr1}) into Eq.~(\ref{chi}) we obtain
the low-temperature limit for the susceptibility in the case
$\alpha <\frac{1}{2}$
\begin{equation}
\chi t=\frac{(1-2\alpha )\left( 1+x\right) ^{2}}{6tx} .
\label{chi1}
\end{equation}

\subsection{The case $\alpha =\frac{1}{2}$}

In the case $\alpha =\frac{1}{2}$ the ground state of $H_{\Delta
}$ is ferromagnetic as well as for the case $\alpha <$
$\frac{1}{2}$. However, the denominator ($1-2\alpha $) presented
in Eqs.~(\ref{Zz1}), (\ref{theta1}), (\ref{corr1}) indicates that
the case $\alpha =\frac{1}{2}$ is special and it is necessary to
keep more terms in the expansion of the energy near the minimum:
\begin{equation}
H_{\Delta }=-\frac{3}{2}+\frac{1}{4}(\theta _{1}-\theta _{2})^{2}-\frac{1}{48%
}(\theta _{1}-\theta _{2})^{4}+\frac{1}{4}\theta _{1}^{2}\theta _{2}^{2}+%
\frac{1}{2}\theta _{1}\theta _{2}(1+\cos \varphi _{2}).  \label{Hh2}
\end{equation}
The form of Hamiltonian (\ref{Hh2}) suggests to change
variables as
\begin{eqnarray}
\theta _{1} &=&ut^{1/4}+vt^{1/2}  \nonumber \\
\theta _{2} &=&ut^{1/4}-vt^{1/2}.  \label{uv}
\end{eqnarray}%
Then to the lowest power in $t$ Hamiltonian (\ref{Hh2}) becomes:
\begin{equation}
\beta H_{\Delta }=-\frac{3\beta }{2}+v^{2}+\frac{1}{4}u^{4}+\sqrt{\beta }%
\frac{1+\cos \varphi _{2}}{2}u^{2}.  \label{Hcrit}
\end{equation}
Substituting Eq.~(\ref{Hcrit}) into the partition function (\ref{Z3})
and integrating it over $\varphi _{2}$ we get
\begin{equation}
Z_{\Delta }=\frac{1}{2}e^{\frac{3}{2}\beta }\beta ^{-5/4}\int_{0}^{\infty
}u^{2}du\int_{0}^{\infty }dv\exp \left( -v^{2}-\frac{1}{4}u^{4}-\frac{1}{2}%
\sqrt{\beta }u^{2}\right) I_{0}\left( \frac{1}{2}\sqrt{\beta }u^{2}\right) .
\label{Zza}
\end{equation}
Now we notice that the argument of the Bessel function in
Eq.~(\ref{Zza}) tends to infinity at $t\to 0$ and we can use the
asymptotic form of the Bessel function
\begin{equation}
I_{0}\left( x\to \infty \right) =\frac{e^{x}}{\sqrt{2\pi x}} .
\end{equation}
Then the partition function can be integrated which yields
\begin{equation}
Z_{\Delta }=\frac{\sqrt{\pi }}{8}\beta ^{-3/2}\exp \left( \frac{3}{2}\beta
\right) .
\end{equation}
The mean value of $\theta _{1}^{2}$ in this case is
\begin{equation}
\left\langle \theta _{1}^{2}\right\rangle =\beta ^{-1/2}\left\langle
u^{2}\right\rangle =\frac{2}{\sqrt{\pi \beta }}
\end{equation}%
which results in the following expressions for the correlators:%
\begin{eqnarray}
\left\langle \vec{n}_{1}\cdot \vec{n}_{2}\right\rangle &=&1-\frac{1}{\sqrt{%
\pi \beta }}  \nonumber \\
\left\langle \vec{n}_{1}\cdot \vec{n}_{3}\right\rangle &=&1-\frac{4}{\sqrt{%
\pi \beta }} . \label{corr2}
\end{eqnarray}
Then, the leading term for susceptibility (\ref{chi}) in the
low-temperature limit in the critical point $\alpha =\frac{1}{2}$
is
\begin{equation}
\chi t=\frac{\sqrt{\pi }\left( 1+x\right) ^{2}}{12xt^{1/2}}.
\label{chi2}
\end{equation}

\subsection{The case $\alpha >\frac{1}{2}$}

As was discussed in Sec.~II the ground state of one triangle in the
region $\alpha >\frac{1}{2}$ is a ferrimagnetic one. Therefore, in
this case we expand the Hamiltonian (\ref{Htri1}) around the
ferrimagnetic classical spin configuration:
\begin{equation}
H_{\Delta }=-\frac{1}{2\alpha }-\alpha +(\alpha -\frac{1%
}{2\alpha })xy+\frac{\alpha }{2}(x^{2}+y^{2})+(2\alpha -%
\frac{1}{2\alpha })z^{2},
\end{equation}%
where variables $x,y,z$ describe the deviations around the ground
state
\begin{eqnarray}
\theta _{1} &=&\theta _{0}+x  \nonumber \\
\theta _{2} &=&\theta _{0}+y  \nonumber \\
\varphi _{2} &=&\pi +2z.
\end{eqnarray}
The partition function in this case can be written in the form
\begin{equation}
Z_{\Delta }=\frac{\sin ^{2}\theta _{0}}{8\pi }\int_{-\infty }^{\infty
}e^{-\beta H_{\Delta }}dxdydz
\end{equation}%
and after integration one obtains%
\begin{equation}
Z_{\Delta }=\beta ^{-3/2}\sqrt{\frac{\pi }{32\alpha }}\exp \left(
\frac{\beta }{2\alpha }+\beta \alpha \right).
\end{equation}
Spin correlators $\left\langle \vec{n}_{1}\cdot
\vec{n}_{2}\right\rangle $ and $\left\langle \vec{n}_{1}\cdot
\vec{n}_{3}\right\rangle $ are given by the ground state
configuration (\ref{theta0}), thermal fluctuations in the region
$\alpha >\frac{1}{2}$ are irrelevant for the calculation of the
leading term in the susceptibility:
\begin{equation}
\chi t=\frac{\left( 2\alpha x+1\right) ^{2}}{6x(4\alpha ^{2}-1)}.
\label{chi3}
\end{equation}

Summarizing our findings for the susceptibility $\chi(t)$ we may
conclude that the power-law divergence as $t \to 0$ is different
in all three regimes $\alpha < \frac{1}{2}$, $\alpha =
\frac{1}{2}$, $\alpha > \frac{1}{2}$, see Eqs.~(\ref{chi1}),
(\ref{chi2}), and (\ref{chi3}).

\section{Scaling near the critical point and finite size effect}

In this section we analyze the obtained analytical results for low
temperatures and estimate the finite-size effect. As follows from
Eqs.~(\ref{corr1}) and (\ref{corr2}) the correlation length (\ref{ksi})
has different low-temperature behavior in different regions:
\begin{eqnarray}
\xi &=&\frac{1-2\alpha }{2t},\qquad \alpha <\frac{1}{2}
\label{xi1} \\
\xi &=&\frac{1}{4}\sqrt{\frac{\pi }{t}},\qquad \alpha =\frac{1}{2}
\label{xi2} \\
\xi &=&\left [\ln \left\vert \frac{2\alpha ^{2}}{1-2\alpha ^{2}}  \right\vert
\right]^{-1},
\qquad \alpha >\frac{1}{2} \;. \label{xi3}
\end{eqnarray}
For $\alpha <\frac{1}{2}$ the correlation length diverges in the
low-temperature limit as $\xi \sim 1/t$ similar to the classical
ferromagnetic chain ($\alpha =0$). In the critical point $\alpha
=\frac{1}{2}$ the correlation length diverges as well, but by
another law $\xi \sim t^{-1/2}$. In the region $\alpha
>\frac{1}{2}$ the correlation length remains finite at $t=0$. This
fact is directly related to the macroscopic degeneracy of the
ground state discussed in Sec.~II, which causes the rapid decay of
the correlation between spins even at zero temperature.

As follows from Eq.~(\ref{xi3}), there are two special cases
$\alpha =1/2$ and $\alpha =1/\sqrt{2}$ where the correlation
length is zero. The case $\alpha =1/2$ correspond to the critical
point, and we will analyze the behavior of the system near the
critical point later.

The case $\alpha =1/\sqrt{2}$ is special, because according to
Eq.~(\ref{theta0}) the angle  $\theta _{0}=\pi /4$, so that
adjacent basal spins in the ground state are orthogonal,
$\vec{n}_{1}\cdot \vec{n}_{3}=0$. This leads to the fact that all
correlators in Eqs.~(\ref{corr}) become zero. However, this does
not mean that all spins at this point become independent. Instead
this is the point where the ferromagnetic type of correlations of
basal spins $\left\langle \vec{n}_{1}\cdot
\vec{n}_{2m+1}\right\rangle \sim \exp (-m/\xi )$ turns into the AF
type: $\left\langle \vec{n}_{1}\cdot \vec{n}_{2m+1}\right\rangle
\sim (-1)^{m}\exp (-m/\xi )$ for $\alpha >1/\sqrt{2}$. So, this is
not a transition point and thermodynamic quantities have no
singularity at $\alpha =1/\sqrt{2}$.

Now let us analyze the correlation function in the ferrimagnetic
region $\alpha >\frac{1}{2}$ at $T=0$. In absence of
thermal fluctuations the system is in the ground state and the
angle between nearest basal spins is fixed and equal to $2\theta
_{0}$ according to Eq.~(\ref{theta0}). However, as was noted in
Sec.~II the spins in each triangle can lie in any plane, which leads to
the degeneracy of the ground state. Each ground state spin
configuration can be represented as a sequence of points lying
on the unit sphere with an equal distance between neighboring
points in the sequence as shown in Fig.~\ref{Fig_sphere}. Thus, the
problem of the spin correlations at zero temperature is equivalent
to the problem of a random walk with fixed finite step length
$2\theta _{0}$ on a unit sphere. Eq.~(\ref{corr}) gives the exact
solution of this problem
\begin{equation}
\cos (2\theta _{m})=\cos ^{m}(2\theta _{0})  \label{walk} \; ,
\end{equation}%
where the angle $\theta _{m}$ is defined by the relation: $\cos
(2\theta _{m})=\left\langle \vec{n}_{1}\cdot
\vec{n}_{2m+1}\right\rangle $. Eq.~(\ref{walk}) is valid for any
value of the step length. In particular, when the step of the random
walk is small $2\theta _{0}\ll 1$ ($\alpha $ is close to the
critical point) and the number of steps is not very large, $\theta
_{m}\lesssim 1$, one can expand both sides of Eq.~(\ref{walk}) and
reproduce the common diffusion law:
\begin{equation}
\theta _{m}=\sqrt{m}\theta _{0}   \; . \label{diffusion}
\end{equation}
Returning to the language of the correlation function, we note
that the condition of the validity of Eq.~(\ref{diffusion}),
$\theta _{m}\lesssim 1$, provides an estimate of the correlation
length $\xi =m$ as $\xi \sim \theta _{0}^{-2}$, which is in accord
with Eq.~(\ref{xi3}) for $\alpha $ close to $\frac{1}{2}$.

\begin{figure}[tbp]
\includegraphics[width=5in,angle=0]{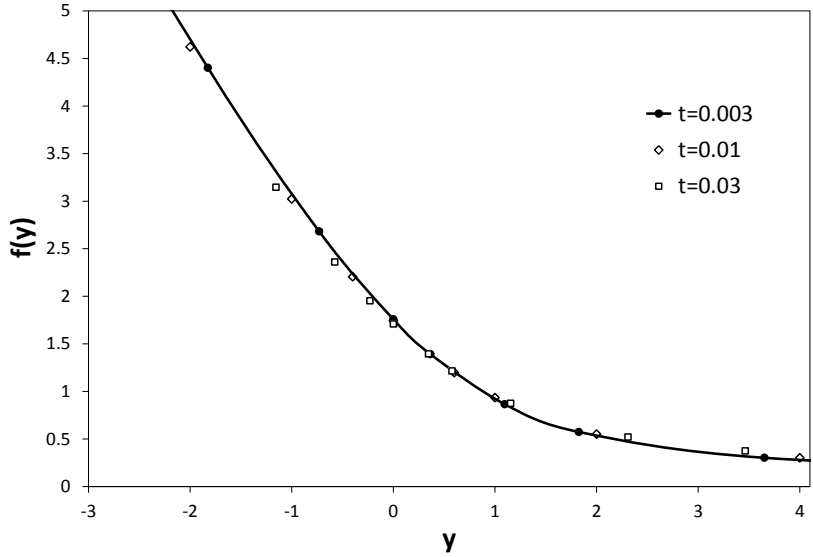}
\caption{Scaling function $f(y)$ for the correlation length and
the susceptibility in the vicinity of the critical point $\alpha
=\frac{1}{2}$ calculated for different values of $\alpha $ and
$t$. All data lie on one curve, which confirms the scaling law.}
\label{Fig_scaling}
\end{figure}

The analysis of Eqs.~(\ref{xi1}), (\ref{xi2}), (\ref{xi3}) in the
vicinity of the point $\alpha =\frac{1}{2}$ allows us to write the
correlation length in the scaling form:
\begin{equation}
\xi (\alpha ,t)=\frac{1}{4\sqrt{t}}f\left( y\right) \; ,
 \label{xi-sc}
\end{equation}%
where the scaling parameter%
\begin{equation}
y=\frac{2\alpha -1}{\sqrt{t}}  \; ,
 \label{scale_x}
\end{equation}%
and the scaling function $f(y)$ has the following asymptotic:%
\begin{eqnarray}
f(y) &\sim &-2y,\qquad y\to -\infty  \nonumber \\
f(y) &\sim &1/y,\qquad y\to \infty  \label{fx}
\end{eqnarray}%
and $f(0)=\sqrt{\pi }$. The scaling function calculated for
different $\alpha $ and $T$ is shown in Fig.~\ref{Fig_scaling}.

Next we note that in the vicinity of the critical point the
correlation length is large, which allows us to use the relation
between the susceptibility and the correlation length
(\ref{chi-xi}):
\begin{equation}
t\chi(\alpha ,t)=\frac{\left( 1+x\right) ^{2}}{12x\sqrt{t}}f(y)  \; .
\label{chi_sc}
\end{equation}
Eq.~(\ref{chi_sc}) describes the scaling form of the susceptibility
in the vicinity of the critical point and correctly reproduces
Eqs.~(\ref{chi1}), (\ref{chi2}), (\ref{chi3}) in the corresponding
limits.

In order to estimate the finite size effect one needs to compare
the chain length with the correlation length, so that another
scaling parameter $N/\xi $ appears. All above results correspond
to the thermodynamic limit when $N\gg \xi $. However, for a finite
chain and low enough temperatures, when $N<\xi $, another regime
takes place. For example, the susceptibility per site in the short
chain case $N<\xi $ as follows from Eq.~(\ref{chi_0}) is
\begin{equation}
t\chi =\frac{(1+x)^2}{6x}N  \; . \label{chi_finite}
\end{equation}
This behavior is obviously a non-thermodynamic one. Here we notice
that Eq.~(\ref{chi_sc}) reduces to Eq.~(\ref{chi-xi}) with
substitution $\xi =N/2$. This allows us to write the scaling form
for the susceptibility in the vicinity of the point $\alpha
=\frac{1}{2}$
\begin{equation}
t\chi (\alpha ,t,N)=\frac{\left( 1+x\right) ^{2}}{3x}\xi (\alpha
,t)F\left( \frac{2\xi (\alpha ,t)}{N}\right)  \label{chi_gen}
\end{equation}%
with $\xi (\alpha ,t)$ given by Eq.~(\ref{xi-sc}). The scaling
function $F(z)$ describes the finite-size effect and has the
limits $F(0)=1$ and $F(y) =1/z$ at $z\to \infty $, smoothly
switching between Eqs.~(\ref{chi-xi}) and (\ref{chi_finite}).
As follows from Eq.~(\ref{chi_gen}) the finite-size effects become
important for low temperatures and $\alpha \leq \frac{1}{2}$, when
the correlation length is large. In the region $\alpha
>\frac{1}{2}$ the correlation length $\xi $ is finite even at zero
temperature and, therefore, the thermodynamics converges rapidly
with $N$.

\section{Quantum effects}

In this section we ascertain a relation of the thermodynamic
properties of the classical model (\ref{H}) with those of the
quantum delta chain (\ref{Hq}). We are mainly interested in the
role of quantum effects in the ferromagnetic phase
($\alpha<\frac{1}{2}$) and especially for the frustration
parameter of the $Fe_{10}Gd_{10}$ system, i.e. $\alpha=0.464$
($a=0.65$). The classical approximation corresponds to the spin
quantum numbers $S\to\infty$. It is known that even for a system
with relatively large values of spins quantum effects become
essential at low temperatures. The analysis of thermodynamic
properties of the quantum model is performed by a combination of
high temperature series expansion (HTE) \cite{11order}, exact
diagonalization (ED) \cite{spinpack} and finite-temperature
Lanczos (FTL) technique \cite{FTL1,FTL2} (where ED and FTL work
only for finite delta chains) as well as by the modified spin-wave
theory (MSWT) \cite{MSWT}. For simplicity we consider next, if not
mentioned otherwise, the delta chain model with $S_{a}=S_{b}=S$.

We start our analysis with the estimate of the quantum corrections
to the classical results for high temperatures. For this aim we
use HTE for the quantum delta chain in powers of $T^{-1}$. Such
expansion has been calculated up to 11th order using the code of
Ref.~\cite{11order}. The first three terms of HTE for the specific
heat are
\begin{eqnarray}
C(S,T) &=&\frac{X^{2}(2+\alpha ^{2})}{3T^{2}}-\frac{4X^{3}\alpha
-X^{2}(2+\alpha ^{3})}{6T^{3}} \nonumber \\
&&-\frac{(3X^{4}+8X^{3}-3X^{2})(2+\alpha ^{4})+5X^{3}(2\alpha +5\alpha ^{2})%
}{45T^{4}}+ O(T^{-5}) \; , \label{CST}
\end{eqnarray}
where $X=S(S+1)$. The HTE for the susceptibility has a similar
structure:
\begin{equation}
\chi(S,T)=\frac{X}{3T}+\frac{X^{2}(2-\alpha)}{9T^{2}}+\frac{X^{3}(2-4\alpha+\alpha^2)
-\frac{3}{4}X^{2}(2+\alpha^2)}{27T^3} \; . \label{ChiST}
\end{equation}
As follows from Eqs.~(\ref{CST}) and (\ref{ChiST}) each HTE term
proportional to $\sim T^{-m}$ contains a factor which is a
polynomial of the order $m$ in $S(S+1)$. The leading terms of
these polynomials are described by the classical approximation
(\ref{Ct_HTE}). Moreover, it turns out that the first term of the
HTE series (\ref{CST}) is exactly reproduced by the classical
expansion (\ref{Ct_HTE}) after the substitution $t=T/S(S+1)$.
Therefore, the leading term of the difference between the
classical and the quantum expansions is of the order of $T^{-3}$:
\begin{equation}
C(S,T)-C_{class}(T) =\frac{X^{2}(2+\alpha ^{3})}{6T^{3}}
+O(T^{-4}) \label{CST2}
\end{equation}
and, similar,
\begin{equation}
\chi(S,T)-\chi_{class}(T) =-\frac{X^{2}(2+\alpha^2)}{36T^3}
+O(T^{-4}) \; . \label{chiST2}
\end{equation}

\begin{figure}[tbp]
\includegraphics[width=5in,angle=0]{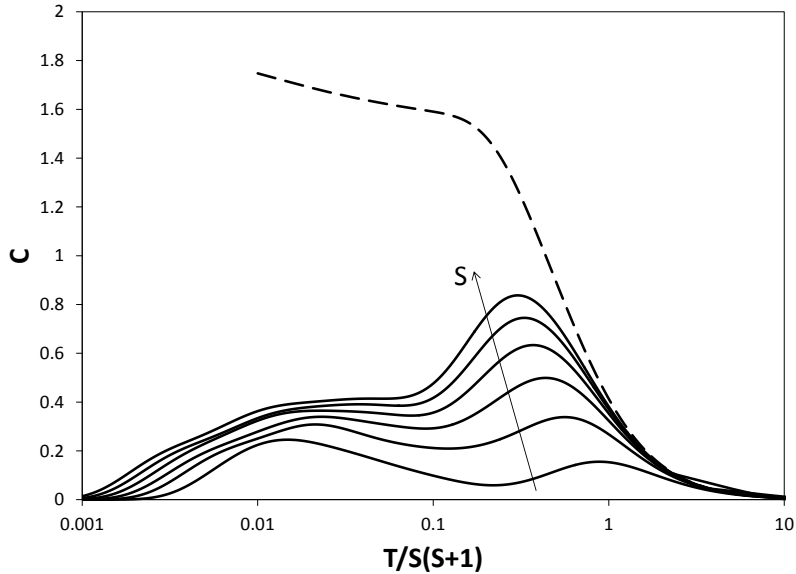}
\caption{Specific heat as a function of the normalized temperature for the quantum
spin model (\ref{Hq}) for $\alpha=0.45$ calculated by ED and FTL
for finite $N=6$ chain. The curves with spin
$S=\frac{1}{2},1,\frac{3}{2},2,\frac{5}{2},3$ are arranged in
order from bottom to top. Specific heat for classical model is
shown by dashed line.} \label{Fig_cTS}
\end{figure}

\begin{figure}[tbp]
\includegraphics[width=5in,angle=0]{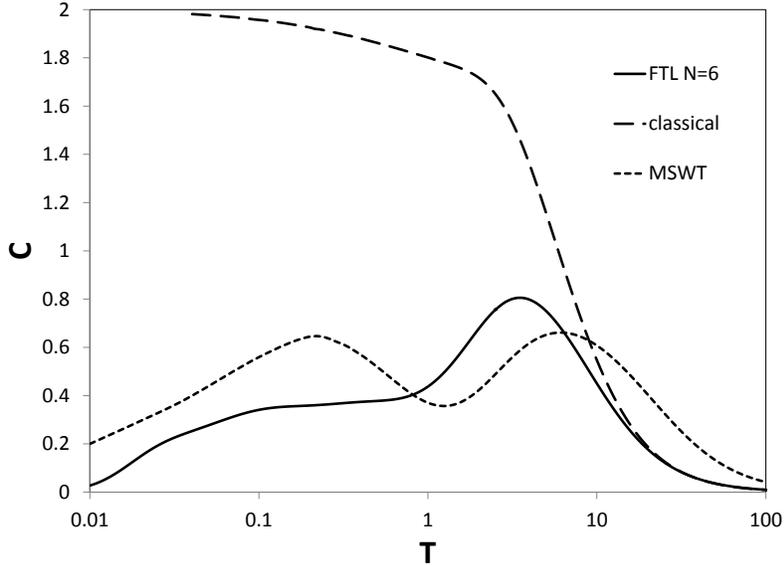}
\caption{Specific heat as a function of temperature for the
$Fe_{10}Gd_{10}$ parameter set  ($a=0.65$, $S_{a}=\frac{7}{2}$,
$S_{b}=\frac{5}{2}$) calculated by the FTL method for $N=6$ (solid
line), in classical approximation (dashed line) and in MSWT
approach (dotted line).} \label{Fig_CT_mol}
\end{figure}

The temperature range where HTE gives reliable results is of the
order of $T/S(S+1) \sim 1$, see also Ref.~\cite{11order}. For the
analysis of the thermodynamic properties of the quantum model at
low temperatures $T/S(S+1) < 1$ we carried out numerical ED and
FTL calculations for finite delta chains. In Fig.~\ref{Fig_cTS} we
present the temperature dependence of the specific heat for a
delta chain of $N=6$ unit cells for different spin values from
$S=\frac{1}{2}$ to $S=3$, where we consider the frustration
parameter $\alpha =0.45$ which is close to the critical point
$\alpha _{c}=\frac{1}{2}$ and is related to the situation in
$Fe_{10}Gd_{10}$. As shown in Fig.~\ref{Fig_cTS} all curves with
different spin coincide with the classical $C(T)$ for $T\geq
S(S+1)$ and deviate from the classical curve and from each other
at $T\sim S(S+1)$, in accord with the prediction of HTE
(\ref{CST2}). It is also seen in Fig.~\ref{Fig_cTS} that for
$S=\frac{1}{2},1,\frac{3}{2}$ the specific heat exhibits a
two-peak structure with maxima at low and intermediate
temperatures. With increasing $S$ the height of the second maximum
increases and its position along the $T/S(S+1)$-axis  shifts to
the left. The low-temperature maximum transforms to a shoulder for
$S\geq 2$. In particular, such shoulder exists (as shown in
Fig.~\ref{Fig_CT_mol}) in the delta-chain with $N=6$ and
$S_{a}=\frac{7}{2},$ $S_{b}=\frac{5}{2}$ ($a=0.65$) modeling the
fictitious `magnetic molecule' $Fe_{6}Gd_{6}$.

Unfortunately, the ED and FTL calculations have the following
limitation: the higher the spin value the shorter the chain that
can be calculated, so that for $S>3$ calculations even for $N=6$
are impossible because the Hilbert space dimension becomes too large. Besides, because of the
finite-size energy gaps in the spectrum the finite chain
calculations cannot correctly describe the low temperature
behavior of the system in the thermodynamic limit $N\to\infty$.
Therefore, a complementary approach is needed to overcome these
shortcomings of finite-chain calculations. We use an approximate
method based on the modified spin wave theory \cite{MSWT}. In this
method the spin operators are replaced by bosonic operators as
in the standard spin-wave theory and the constraint of zero total
magnetization at finite temperature is imposed. This approach has
been successfully applied to low-dimensional Heisenberg models
\cite{MSWT,Yamamoto,Sirker}. Retaining only the lowest order in
the spin-bosonic transformation terms we obtain a bilinear bosonic
Hamiltonian, which after the diagonalization takes a form
\begin{equation}
\hat{H}=\sum \varepsilon _{A}(k)A_{k}^{\dagger }A_{k}+\varepsilon
_{B}(k)B_{k}^{\dagger }B_{k} \; ,
\end{equation}%
where two magnon branches $\varepsilon _{A,B}(k)$ are%
\begin{equation}
\varepsilon _{A,B}(k)=2S(1-\alpha \sin ^{2}\frac{k}{2})\pm
2S\sqrt{(1-\alpha \sin ^{2}\frac{k}{2})^{2}-(1-2\alpha )\sin
^{2}\frac{k}{2}} \; .
\end{equation}
The energy is%
\begin{equation}
E=\sum [\varepsilon _{A}(k)n_{A}(k)+\varepsilon _{B}(k)n_{B}(k)] \; .
\end{equation}
Here the mean values of the occupations $n_{A}(k)=\left\langle
A_{k}^{\dagger }A_{k}\right\rangle $ and $n_{B}(k)=\left\langle
B_{k}^{\dagger }B_{k}\right\rangle $ are given by the
Bose-Einstein distribution:
\begin{equation}
n_{A,B}(k)=\left[ \exp \left( \frac{\varepsilon _{A,B}(k)-\mu }{T}\right) -1%
\right] ^{-1}
\end{equation}%
and the chemical potential $\mu $ is defined by the condition of
zero total magnetization:
\begin{equation}
\frac{1}{N}\sum [n_{A}(k)+n_{B}(k)]=2S \; .
\end{equation}

\begin{figure}[tbp]
\includegraphics[width=5in,angle=0]{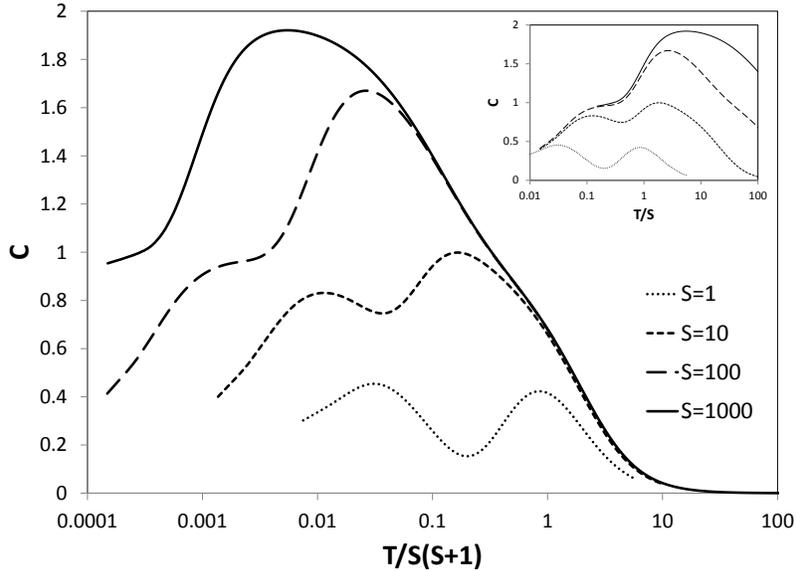}
\caption{Specific heat as a function of the normalized temperature
for the  quantum spin model (\ref{Hq}) for $\alpha=0.45$ in
MSWT approach for $S=1,10,100,1000$. In the inset the same curves
are plotted vs. $T/S$.} \label{Fig_MSWT_S}
\end{figure}

The results of the MSWT calculations of the specific heat for
$\alpha =0.45$ are presented in Fig.~\ref{Fig_MSWT_S}. As can be
seen from Fig.~\ref{Fig_MSWT_S} the specific heat for $S=1$
exhibits  a double-peak structure similar to that in
Fig.~\ref{Fig_cTS}. Such a structure is related to the fact that
the lower $\varepsilon_{A}$ and the higher $\varepsilon_{B}$
magnon branches give distinct contributions to $C(T)$ at low and
intermediate temperatures because these branches are well
separated for $\alpha $ close to $\frac{1}{2}$. In this case the
gap between the branches is $\Delta E=4S\alpha$. At low
temperatures $C(T)$ is determined by the lower branch which
behaves at $k\to 0$ as
\begin{equation}
\varepsilon _{A}(k)=\frac{S(1-2\alpha )}{4}k^{2} \; .
\end{equation}
The lower branch sets the low-energy scale leading to a
low-temperature peak in the specific heat. The low-temperature
maximum exists even for $S=10$ contrary to the results of the ED
and the FTL calculations and this maximum transforms to the
           shoulder for $S\sim 100$ only. This means that the MSWT
approximation overestimates the value of $C(T)$ at low
temperature. The second maximum increases with growing
$S$ and tends to the classical value $C=2$ as
$S\to\infty$. Its position shifts to the left of
the $T/S(S+1)$-axis with increasing $S$, which is in
agreement with ED and FTL results. Such behavior is a
consequence of the fact that in the MSWT approximation the
temperature $T$ and the spin value $S$ form a scaling variable
$T/S$ at low temperatures (because $\varepsilon _{A,B}\sim S$ and
$\mu\sim T^2/S^3$ give negligible contributions at low
$T$). Therefore, the low temperature behavior of the delta
chain with high value of $S$ is a function of $T/S$ and,
therefore, the shoulder and the maximum in the specific heat are
located at $T\sim S$ as illustrated in the inset of
Fig.~\ref{Fig_MSWT_S}.

The specific heat of the quantum model tends to zero in the limit
$T\to 0$. It is believed that the spin-wave approach gives the true
leading term in this limit:
\begin{equation}
C(T)=\frac{3\zeta (\frac{3}{2})}{4\sqrt{\pi
}}\sqrt{\frac{T}{(1-2\alpha)S}} \; .
\end{equation}%
This result is related to the infinite delta-chain. For finite
systems such as $Fe_{10}Gd_{10}$ the specific heat vanishes
exponentially in $1/T$ at $T\to 0$. In contrast to the quantum
case, the classical specific heat is finite at $T=0$. Therefore,
the behavior of the specific heat at low temperatures of the
classical and the quantum delta chain is essentially different.
\begin{figure}[tbp]
\includegraphics[width=5in,angle=0]{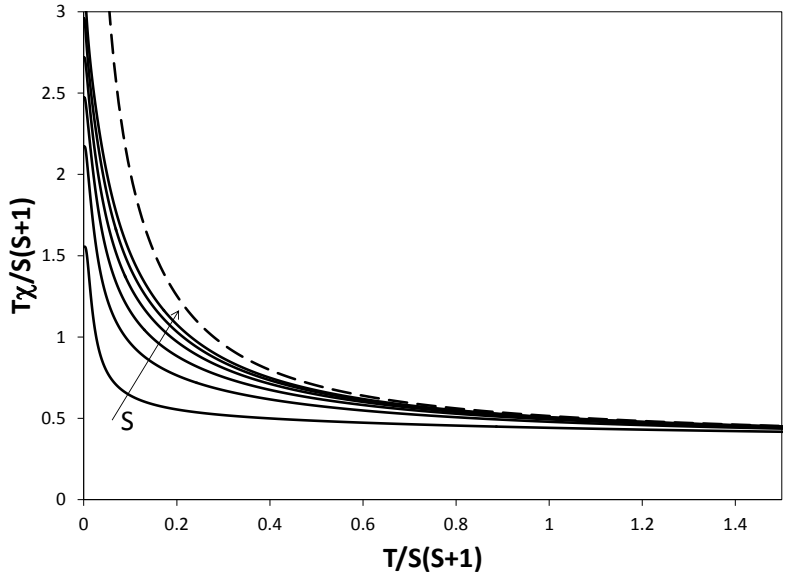}
\caption{Susceptibility as a function of the normalized temperature for the
quantum
spin model (\ref{Hq}) for $\alpha=0.45$ calculated by ED and FTL
for finite $N=6$ chain. The curves with spin
$S=\frac{1}{2},1,\frac{3}{2},2,\frac{5}{2},3$ are arranged in
order from bottom to top. The susceptibility for the classical model is
shown by dashed line.} \label{Fig_chiTS}
\end{figure}
However, in some cases the low-temperature properties of the
quantum and the classical models are very similar. For example,
both classical and quantum ferromagnetic chains have the same
universal magnetic low-temperature behavior
\cite{Nakamura,Sachdev,Bacalis}. Similarly, such universality
holds for the F-AF chain as well \cite{DK11}. In analogy to these
models we can expect that the low-temperature behavior of the
zero-field susceptibility $\chi (T)$ of the classical and the
quantum delta-chain in the ferromagnetic phase are also the same.
The dependencies of the product $\chi T$ of the classical delta
chain and the quantum model with $N=6$ and different $S$ are shown
in Fig.~\ref{Fig_chiTS} for $\alpha =0.45$. With increasing $S$
the quantum curves approach the classical one. The low-temperature
susceptibility in the MSWT
approximation is%
\begin{equation}
\chi =\frac{4(1-2\alpha )S^4}{3T^2}-\frac{2\zeta
(\frac{1}{2})(1-2\alpha )^{1/2}S^{5/2}}{\pi ^{1/2}T^{3/2}}+\ldots \; \; .
\label{chi_MSWT}
\end{equation}
The comparison of Eqs.~(\ref{chi_MSWT}) and (\ref{chi1}) shows
that the leading terms of the classical and the quantum
susceptibility at $T\to 0$ coincide. We note also that the leading
low-temperature term for the correlation length in quantum model
coincides with the classical result (\ref{xi1}),
$\xi=(\frac{1}{2}-\alpha)S^2/T$. According to Eq.(\ref{chi_MSWT})
the susceptibility $\chi (T)$ for infinite system diverges as
$T^{-2}$. However, for finite systems and very low temperatures,
when $(\frac{1}{2}-\alpha)S^2>NT$, $\chi(T)$ diverges as $T^{-1}$.
It is due to the fact that according to Eq.~(\ref{xi1}) the
correlation length at
$T\to 0$ exceeds the system size and the product $\chi T$ at $T=0$ is%
\begin{eqnarray}
\chi T &=&\frac{4}{3}N^{2}S^{2}  \nonumber \\
\chi T &=&\frac{4}{3}N^{2}S^{2}(1+\frac{1}{2NS})
\label{chi_q_finite}
\end{eqnarray}%
for the classical and the quantum model, respectively. The
term in parenthesis in Eq.~(\ref{chi_q_finite}) is the quantum
correction to the classical result.

\begin{figure}[tbp]
\includegraphics[width=5in,angle=0]{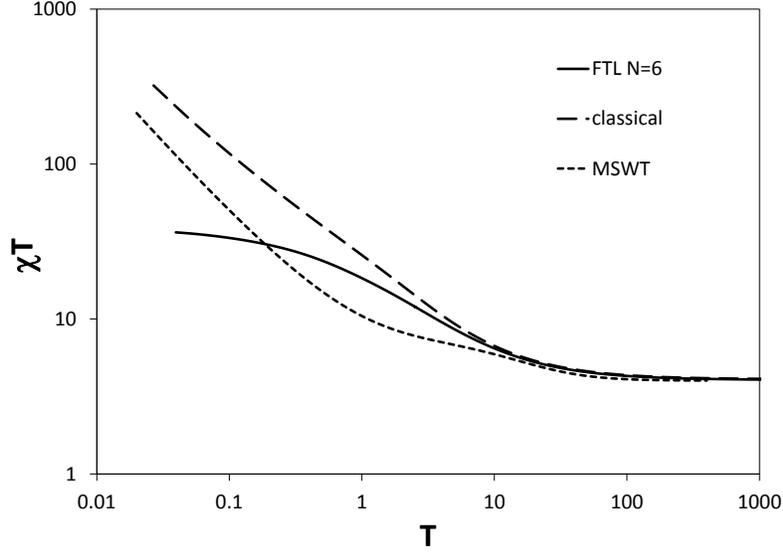}
\caption{Susceptibility as a function of temperature for the
$Fe_{10}Gd_{10}$ parameter set
($a=0.65$, $S_{a}=\frac{7}{2}$,
$S_{b}=\frac{5}{2}$) calculated by the FTL method for $N=6$ (solid
line), in classical approximation (dashed line) and in MSWT
approach (dotted line).} \label{Fig_chiT_log_mol}
\end{figure}

According to Eq.~(\ref{chi_q_finite}) $\chi T$ tends to finite
value at $T=0$ for finite systems including real compound
containing molecules $Fe_{10}Gd_{10}$ and such behavior is
visualized in Fig.~\ref{Fig_chiTS} and Fig.~\ref{Fig_chiT_log_mol}.

\section{Summary}

In this paper we have studied the delta-chain with competing
ferro- and antiferromagnetic interactions. This model finds its
finite-size material realization in the recently synthesized
cyclic compound $Fe_{10}Gd_{10}$  \cite{S60}. In dependence  on
the frustration parameter $\alpha$ it exhibits ferromagnetic and
ferrimagnetic   ground-state phases separated by a critical point
at  $\alpha=1/2$, where the ground state of the quantum model at
this point exhibits a massive degeneracy. For the classical
version of this model, which seems to provide a reasonable
description $Fe_{10}Gd_{10}$ down to moderate temperatures, we
obtain exact results for the partition function and the
thermodynamics. The explicit analytical expansion of thermodynamic
quantities is provided for high temperatures. The low temperature
behavior of the specific heat and the susceptibility of the
classical model is different in various phases. In the
ferromagnetic phase ($\alpha <\frac{1}{2}$) $C(T)=2$ at $T=0$,
while $C(0)=\frac{3}{2}$ in the ferrimagnetic phase ($\alpha
>\frac{1}{2}$). The zero-field susceptibility diverges as $T^{-2}$
and $T^{-1}$ for $\alpha <\frac{1}{2}$ and $\alpha >\frac{1}{2}$,
respectively. In the critical point $\alpha =\frac{1}{2}$ the
susceptibility behaves as $\chi \sim T^{-3/2}$.

The classical model corresponds to the limit $S\to\infty $.
Quantum corrections to the classical results for large but finite
$S$ are small at high temperature ($T>S^{2}$). However, the
quantum effects become essential at low temperature. In
particular, the classical specific heat is finite at $T=0$, while
$C(T)\sim \sqrt{T/S(1-2\alpha )}$ in the infinite quantum model
for $\alpha <\frac{1}{2}$ and it is exponentially small at $T\to
0$ for finite delta-chain such as $Fe_{10}Gd_{10}$ magnetic molecule.
On the other hand,
the leading term of the susceptibility of the classical and
the quantum models coincide for both small and large temperatures
at $\alpha <\frac{1}{2}$. The product $\chi T$ diverges as
$T^{-1}$ at $T\to 0$ for $\alpha <\frac{1}{2}$ in the infinite
chain and it is proportional to $N^{2}$ for finite systems. Such
behavior of $\chi T$ takes place, in particular, in the
$Fe_{10}Gd_{10}$ molecule.

The $C(T)$ dependence in the ferromagnetic phase is characterized
by the existence of a two-peak structure for $S<2$. For  $S\geq 2$
the low-temperature maximum transforms to the shoulder. Such
a $C(T)$ profile with a shoulder and a maximum was observed
for the magnetic part of $C(T)$ of $Fe_{10}Gd_{10}$
\cite{S60}. The shoulder and the maximum shift towards higher
temperature as $S$ increases and $T_{\max }\sim S$.

\end{document}